------------------------------------------------------------------------------
\documentstyle[preprint,aps]{revtex}
\begin{document}

\draft \title{A Generalized Circle Theorem on Zeros of Partition Function at
Asymmetric First Order Transitions}

\author{ Koo-Chul Lee }

\address{Department of physics and the Center for Theoretical Physics  \\
         Seoul National University \\ Seoul, 151-742, Korea} \date{\today}
\maketitle

\begin{abstract}

We present a generalized circle theorem which includes the Lee-Yang theorem for
symmetric transitions as a
special case.  It is found that zeros of the partition
function can be written in terms of discontinuities in the derivatives of the
free energy.  For asymmetric transitions, the locus of the zeros is tangent to
the
unit circle at the positive real axis in the thermodynamic limit. For
finite-size
systems, they lie off the unit circle if the partition functions of the two
phases
are
added up with unequal prefactors.  This conclusion is substantiated by explicit
calculation of zeros of the partition function for the Blume-Capel model near
and at the triple line at low temperatures.

 \end{abstract}

\pacs{05.50.+q, 05.70.Fh, 64.60.-i, 75.10.Hk}

\narrowtext More than three decades have passed since Yang and Lee
\cite{leeyang52} first published their celebrated papers on the theory of phase
transitions and the circle theorem on the zeros of the partition function of
ferromagnetic Ising models.  Although there have been some developments in
extending the theorem \cite{griffiths73}, little is known about what happens to
the theorem in the case of more general first-order transitions such as
asymmetric transitions or temperature driven transitions.  In their original
proof of the theorem, Lee and Yang relied on particular properties of the
coefficients of the polynomial.  Furthermore they used an analogue of a certain
electrostatic problem to relate the discontinuity in the spontaneous
magnetization to the density of zeroes.  In this paper we present a simple
proof
of the circle theorem which can be extended to first-order phase transitions of
the most general kind.  The extended theorem can be used to (1) identify the
order of the phase transition, since a second-order phase transition manifests
drastically different characteristics in the complex partition function
\cite{leek2}; (2) pinpoint the transition point very accurately; and (3)
resolve the recent controversy over the equal weight versus equal height of the
probability distribution function at an asymmetric transition
\onlinecite{bind1,bind2,borg90} among other potential applications.

The generalized theorem is based on a very simple property that the probability
distribution for the {\it order parameter} $M$, conjugate to an external
ordering field  $H$ is doubly peaked in the two-phase
region.  This simple property alone is sufficient to prove the generalized
circle theorem which states that the locus of zeros of partition function in
the
complex  $e^H$-plane forms a circle near the positive real axis. Near a triple
line where three phases coexist, there are three
peaks in the distribution function, and this leads to some interesting behavior
of loci of complex zeros.  These conclusions are substantiated by explicitly
calculating zeros of the partition function of the Blume-Capel model near and
at
the triple line at low temperatures.

Let us consider a partition function $Z(H)$ of a discrete lattice system made
of
 a $d$-dimensional cube of side $L$ as a
function of variable $H$; we seek the complex zeros of $Z(H)$. $H$ can be an
external magnetic field for a spin system in which
case the conjugate {\it order parameter} would be the magnetization.

In the following we will consider only three phase coexistence for the sake of
simplicity.  Generalization to coexistence of more than three phases is
straightforward if cumbersome.  If we label the three phases as $A, B$ and $C$
and the partition function of each phase as $Z_{\lambda}(H)$, where $\lambda =
A,B,C$, then we have near a triple point \begin{equation}
Z(H)=\sum_\lambda{Z_{\lambda}(H)}.\label{eq:part} \end{equation}

 The fact that the partition function is made of three parts follows from the
fact that there are two barriers, say at $K_1$ and $K_2$ in the order parameter
$M$, which separate three phases.  Therefore we can define
$Z_A(h)=\sum_{M=M_{min}}^{M=K_1}e^{hM}Z_M$,
$Z_B(h)=\sum_{M>K_1}^{M=K_2}e^{hM}Z_M$ and
$Z_C(h)=\sum_{M>K_2}^{M=M_{max}}e^{hM}Z_M$, with $h=H/K_BT \equiv \beta H$.
$Z_M$ is a partition function for fixed $M$ in case $H$ is the external field
and
we may take $\Omega(E)$, the number of configurations at energy $E$, when $H=T$
and $M=E$ for temperature driven transitions.  We can always make $h=0$ at the
coexistence point by redefining $Z_M$ appropriately.

Let $\bar{M}$ be $\max(|M_{min}|, |M_{max}|)$.  Introducing the probability
distribution function $p(x)=Z_{M=x\bar{M}}/Z(H=0)$, we can define the moment
generating function ${\cal M}(t) = \langle e^{tx}\rangle \equiv \sum_x
e^{tx}p(x)
= \sum_{l=0}^{\infty}\langle x^l\rangle t^l/l!$.
Since $\|\langle x^l\rangle \| \leq 1$,  ${\cal M}(t)$ is analytic in the whole
complex $t$-plane.  Since ${\cal M}(t=h\bar{M})
= Z(h)/Z(0)$, zeros of partition function $Z(h)$ can be obtained from zeros of
${\cal M}(t)$.  By separating the contribution from three phases, we can write
${\cal M}(t) = {\cal M}_A(t) +{\cal M}_B(t)+{\cal M}_C(t)$,
where ${\cal M}_{\lambda}(t) = Z_{\lambda}(h=t/\bar{M})/Z(0)$.

We first consider the case where only two phases $A$ and $C$ coexist.
Introducing the cumulant generating function defined by $\psi_{\lambda}(t)
=\ln({\cal M}_{ \lambda}(t))$, we can rewrite
${\cal M}(t) = 2e^{\bar{\psi}(t)}\cosh(\tilde{\psi}(t))$,
where $\bar{\psi} = (\psi_C + \psi_A)/2$ and
$\tilde{\psi} = (\psi_C - \psi_A)/2$.  Zeros of the
${\cal M}(t)$ in the complex $t$-plane are simply solutions of following
equation,
$\tilde{\psi}(t_k) = \pm i(1/2+k)\pi$, where $k=0,1,2 ,3\cdots$.  Zeros of
partition function in the complex $h_k$ can be calculated from the relation
$h_k=t_k/\bar{M}$.

In order to express the zeros of $Z(h)$, in terms of the free energy density
and its derivatives with respect to $h$ of each phase, we expand
$\psi_{\lambda}$
in
Taylor series of $h$ as $\psi_{\lambda} = \ln({\cal M}_{\lambda}(0)) + L^d\sum
_{l=1}^{\infty}h^l\gamma_l^{\lambda}/l!$,
where $\gamma_l^{\lambda}=L^{-d}\partial^l \ln(Z_{\lambda}(h))/\partial
h^l\big|_{h=0}$ .
Therefore $\tilde{\psi}(t=h\bar{M}) = -\ln(a)/2 + \tilde{F}(h)$,
where $a={\cal M}_A(0)/{\cal M}_C(0)$ and $\tilde{F}(h) =
L^d\sum_{l=1}^{\infty}h^l\tilde{\gamma}_l/l!$ with  $\tilde{\gamma}_l
= ( \gamma_l^C-\gamma_l^A)/2$.

Therefore zeros of $Z(h)$ are solutions of the following equations,

\begin{eqnarray}
\Im(\tilde{F}(h)) = \pm(1/2 + k)\pi \equiv I_k \label{eq:zeroi}\\
\Re(\tilde{F}(h)) = \ln(a)/2 \equiv R.  \label{eq:zeror}
\end{eqnarray}

We can easily obtain zeros $Z(h)$ by
inverting the series. Following the tradition, we will consider zeros in the
$z$-plane defined by $z=e^h$. The modulus $r=|z| =e^{\Re(h)}$ and argument
 $\theta = \ln(z/r)/i = Im(h)$ of zeros can be found from

\begin{eqnarray}
\lefteqn{\ln(r_k)=\Re(h_k) =\hat{R} + \hat{\gamma}_2(-\hat{R}^2
+ \hat{I}_k^2) + (2\hat{\gamma}_2^2- \hat{\gamma}_3)
}\hspace{10.0cm} \nonumber \\
\lefteqn{\times(\hat{R}^3 - 3\hat{R}\hat{I}_k^2)
+ (-5\hat{\gamma}_2^3 + 5\hat{\gamma}_2  \hat{\gamma}_3
- \hat{\gamma}_4)
(\hat{R}^4 - 6\hat{R}^2\hat{I}_k^2 + \hat{I}_k^4)
}\hspace{10.0cm} \nonumber \\
\lefteqn{+ (14\hat{\gamma}_2^4-21\hat{\gamma}_2^2\hat{\gamma}_3
+ 3\hat{\gamma}_3^2
+ 6\hat{\gamma}_2\hat{\gamma}_4
- \hat{\gamma}_5)
(\hat{R}^5 - 10\hat{R}^3\hat{I}_k^2}\hspace{10.0cm} \nonumber \\
\lefteqn{ + 5\hat{R}\hat{I}_k^4)
+ (42\hat{\gamma}_2^5 -84\hat{\gamma}_2^3\hat{\gamma}_3
+ 28\hat{\gamma}_2\hat{\gamma}_3^2
+28\hat{\gamma}_2^2\hat{\gamma}_4
- 7\hat{\gamma}_3\hat{\gamma}_4
}\hspace{10.0cm} \nonumber \\
\lefteqn{- 7\hat{\gamma}_2\hat{\gamma}_5
+ \hat{\gamma}_6)
(-\hat{R}^6
+ 15\hat{R}^4\hat{I}_k^2
- 15\hat{R}^2\hat{I}_k^4
+ \hat{I}_k^6)\cdots}\hspace{10.0cm}
\label{eq:realzero} \end{eqnarray}

\begin{eqnarray}
\lefteqn{\theta_k=\Im(h_k) = \hat{I}_k -
2\hat{\gamma}_2\hat{R}\hat{I}_k
+(2\hat{\gamma}_2^2- \hat{\gamma}_3)(3\hat{R}^2\hat{I}_k-\hat{I}_k^3)
}\hspace{10.0cm} \nonumber \\
\lefteqn{+ 4(5\hat{\gamma}_2^3
-5\hat{\gamma}_2\hat{\gamma}_3 +
\hat{\gamma}_4)(-\hat{R}^3\hat{I}_k+\hat{R}\hat{I}_k^3)
+(14\hat{\gamma}_2^4
-21\hat{\gamma}_2^2\hat{\gamma}_3}\hspace{10.0cm} \nonumber \\
\lefteqn{ + 3\hat{\gamma}_3^2
+6\hat{\gamma}_2\hat{\gamma}_4-\hat{\gamma}_5)(5\hat{R}^4\hat{I}_k -
10\hat{R}^2\hat{I}_k^3 +\hat{I}_k^5)
}\hspace{10.0cm} \nonumber \\
\lefteqn{+(-42\hat{\gamma}_2^5
+84\hat{\gamma}_2^3\hat{\gamma}_3 - 28\hat{\gamma}_2\hat{\gamma}_3^2
-28\hat{\gamma}_2^2\hat{\gamma}_4 + 7\hat{\gamma}_3\hat{\gamma}_4 +
7\hat{\gamma}_2\hat{\gamma}_5}\hspace{10.0cm} \nonumber \\
\lefteqn{ - \hat{\gamma}_6)(6\hat{R}^5\hat{I}_k -
20\hat{R}^3\hat{I}_k^3 +6\hat{R}\hat{I}_k^5)\cdots}\hspace{10.0cm}
\label{eq:imaginaryzero} \end{eqnarray}

 where $\hat{R}\equiv R/L^d\mu, \hat{I}_k\equiv I_k/L^d\mu$,
$\hat{\gamma}_l\equiv
\tilde{\gamma}_l/\mu l!$ and $\mu\equiv \tilde{\gamma}_1$, is one half of the
discontinuity of the order parameter across the phase boundary.

The angular density of the zeros, $g(\theta)$ defined by
$L^dg(\theta)=1/(\theta_{k+1}-\theta_k)$, can be obtained by differentiating
the
left hand side of (\ref{eq:zeroi}) with respect to $\theta_k$.  We have
\begin{equation} 2\pi g(\theta) = 2\tilde{\gamma}_1-\tilde{\gamma}_3\theta^2
+\tilde{\gamma}_5\theta^4/12-\cdots + O(L^{-d}) \label{eq:density}
\end{equation}
 In the scaling limit, we have
$r=\exp[\ln(a)/(2L^d\mu)+O(L^{-2d})]$ and
$\theta_k=\pm\pi(1/2+k)/(L^d\mu)+O(L^{-2d})$.  This implies that the zeros are
distributed uniformly on a circle of varying radius near the positive real axis
for finite system if
the partition functions of the two phases are added up with unequal prefactors.
If the prefactors are equal, {\it i.e}, $a=1$, then zeros form a circle of unit
radius near the positive real axis in the scaling limit. However if the
argument
, $\theta_k$ grows large enough zeros deviate from the unit circle unless  all
$\tilde{\gamma}_l$'s with  even $l$ vanish.
However for symmetric transitions we have not only $a=1$ but also
$\tilde{\gamma}_l=0$ for all even $l$.  Therefore we have  $r_k=1$
for all sizes and the angular distribution of the zeros becomes exactly Eq.
(\ref{eq:density}) without the size dependent correction terms.  This is the
Lee-Yang theorem.  We now have the angular density of zeros in a complete power
series in $\theta$.

Let us now consider a case where three phases coexist. Using $F_B(h)$ and $b$
defined by $F_B(h) = L^d\sum_{l=1}^{\infty}h^l\gamma_l^B/l!$ and  $b={\cal
M}_B(0)
/{\cal M}_C(0)$, we can modify eqs.(\ref{eq:zeroi}) and (\ref{eq:zeror}) as

\begin{eqnarray}
e^{\Re(\bar{F}(h))}(e^{\Re(\tilde{F}(h))}
+ae^{-\Re(\tilde{F}(h))})\cos(\Im\tilde{F}
(h)) \nonumber \\=-be^{\Re(F_B(h))}\cos(\Im(F_B(h))), \label{eq:threezeror} \\
e^{\Re(\bar{F}(h))}(e^{\Re(\tilde{F}(h))}
-ae^{-\Re(\tilde{F}(h))})\sin(\Im(\tilde{F}
(h)) \nonumber \\= -be^{\Re(F_B(h))}\sin(\Im(F_B(h))),  \label{eq:threezeroi}
\end{eqnarray}

where $\bar{F}(h)=\bar{\psi}(t=h\bar{M})$.
General solution to the above equation is complicated if not impossible to
obtain.
Therefore we will consider a special case which is useful in analyzing the
ensuing example.  We will consider the case where $a$ and $c$ phases are
symmetric and $b$ phase is also symmetric with respect to the midpoint between
the two.  Furthermore we will consider the scaling limit where corrections in
$O(L^{-2d})$ is neglected.  In that case we have \begin{eqnarray}
\cosh(\Re(\tilde{F}(h)))\cos(\Im(\tilde{F}(h)))= -b/2, \label{eq:tzeror} \\
\sinh(\Re(\tilde{F}(h)))\sin(\Im(\tilde{F}(h)))= 0.  \label{eq:tzeroi}
\end{eqnarray} If $b/2\leq 1$, we have zeros as $\Im(h) =
\cos^{-1}(-b/2)/L^d\mu$ and $\Re(h)=0$ while for $b/2>1$, $\Im(h) =
(\pi(1+2k))/L^d\mu$ and $\Re(h)=\pm\mid\cosh^{-1}b/2\mid/L^d\mu$.  This means
that crossing the triple point from two phase region($b/2\leq 1$) to the single
phase region($b/2>1$), zeros on the unit circle disappear.  Only for small
sized
systems and close to the triple point, we see two circles separating out from
the unit circle which are sort of {\it ghost} circles that separate three
phases, two unstable symmetric and the new stable phases.

We have tested the above  conclusion explicitly by calculating  zeros of the
partition
function of the Blume-Capel model. Consider the energy of the Blume-Capel
model\onlinecite{blume66,capel66}, $E= -H\sum_{i}S_i + D\sum_{i}S_i^2 -J
\sum_{<i,j>} S_iS_j$ of $N$ spins on a square lattice of side $L$($N=L^2$) on a
torus. $H, D$ and $J$ are usual external parameters and $S_i$ are spin
variables
which take three values, $\pm 1$ and $0$. If we designate by $N_+$ and $N_-$
the number of $+1$ spins and $-1$ spins, and $N_e\equiv [(q/2)\sum_{i}S_i^2 -
\sum_{<i,j>} S_iS_j]$, we can write the energy $E=-H(N_+ - N_-) + \Delta((N_+ +
N_-) + JN_e$ where $q$ is the coordination number and $\Delta\equiv  D-Jq/2$.
We
have defined {\it energy} $N_e$ such a way to make the ground state a triple
point,
in the absence of external fields $H$ and $\Delta$. The partition function can
be calculated according to
\begin{equation}
Z(T,H,\Delta)=\sum_{N_e,N_+,N_-}e^{-E(N_e,N_+,N_-)/k_BT}\omega(N_e,N_+,N_-),
\label{eq:partbc}
\end{equation}
where $\omega(N_e,N_+,N_-)$ is the number of configurations of given $N_e,N_+$
and $N_-$. We have calculated exact $\omega(N_e,N_+,N_-)$'s for all $N_+,N_-$
for
$N_e = 0,1,\cdots,6$.
These $\omega$'s are sufficient to calculate the partition
functions at temperatures up to $k_BT/J=0.2$ for sizes up to
$L=70$ with the machine precision.  The phase diagram in $T-H-\Delta$ space
looks like Fig.~\ref{fig1}\cite{Griffiths70}.

Defining order parameters, $\dot{N}=N_++N_-$ and $M=N_+-N_-$ and their
conjugate
field variables, $\zeta=\beta(\Delta-\Delta_o)$ and $\eta=\beta (H-H_o)$, we
can
write the
partition function as  $Z(\zeta,\eta)=\sum_{\dot{N},M}e^{-\zeta\dot{N}+\eta
M}Z_o(\dot{N},M)$.  In the above $H_o$ and $\Delta_o$  are the
values used to calculate the {\it microcanonical}  partition function
$Z_o(\dot{N},M)$ which is
defined as $Z_o(\dot{N},M)=e^{-\beta \Delta_o \dot{N}+\beta H_o
M}\sum_{N_e}e^{-\beta JN_e}\omega(N_e,(\dot{N}+M)/2,(\dot{N}-M)/2)$. If $H_o$
and
$\Delta_o$ are chosen to be the transition point,  $\zeta$ and $\eta$ will
vanish at the transition point.

In order to maximize the finite size
effect, we demonstrate zeros at the highest allowable
temperature, $\beta =5.0$ where  the high-order cumulants are appreciable while
the contributions to $Z_o(\dot{N},M)$ from terms
of $N_e$ higher than $6$ are still negligible. The transition points are
determined by equating the free energies of the coexisting phases for the
system
size $L \simeq 60$. We found the triple point at this temperature,
$P_T=(\zeta_o\equiv \beta\Delta_o=-0.0000454657,
\eta_o\equiv \beta H_o=0)$(Fig.~1(e)). A typical asymmetric transition point
which  we chose for the demonstration is
$P_A=(\zeta_o=1,\eta_o=1.0000061392784119)$, a point like Fig.~1(f).

In Fig.~2. we display $\ln(Z_o(N_++N_-,N_+-N_-)$ by the size of the filled
circle in
($N_+,N_-$) plane at various points in the phase diagram.   It should be noted
that in these figures $M$ and $\dot{N}$ vary along the two diagonal lines.

We found that zeros in the complex $z_{\eta}(\equiv e^\eta)$-plane at phase
points
like Fig.~1(a) along the $\eta_o=0$ line up to the triple point are distributed
more
or less uniformly on the unit circle for all $L$ as predicted by the theorem.
On
the other hand there are no zeros along the zero-field line in
$z_{\zeta}(\equiv
e^\zeta)$-plane in the direction parallel to $\eta_o=0$ line as long as the
phase
points remain far away from the triple point.  At $P_T$ we found zeros in
$z_{\zeta}$-plane in the direction indicated by Fig.~1(e).  form a circle of
varying radius which is dependent on the size of the system as shown in
Fig.~3A.
Since the transition at $P_T$ in the direction parallel to $\eta_o=0$ line is
asymmetric, it might be concluded that the partition functions of the two
phases
are added up with unequal prefactors.  However the asymmetry in this case is
due
to the fact that it is a transition from the two-phase region to the single
phase
region, and the asymmetry factor $a$ in Eq.(\ref{eq:zeror}) is indeed exactly
$2$
if we fit the concentric circles by the predicted relations,
$r=\exp[\ln(a)/(2L^d\mu)]$.

On the other hand in the $z_{\eta}$-plane the uniformly
distributed  $2N$ zeros pair up and converge to each other as $\zeta_o$
approach
the triple point and  they eventually  double up before they bifurcate into two
sets of concentric circles of $N$ zeros each at $P_T$.   Fig.~3B  is the plot
of
zeros at $P_T^{\prime}=(\zeta_o = 0.01088, \eta_o=0)$, a point  slightly to the
right of  Fig.~1(b).  The size dependence in this plot is not from the
$L$'s in $\tilde{F}(h)$ in Eqs.(\ref{eq:tzeror}) and (\ref{eq:tzeroi})  which
would give
converging circles rather than divergence.  The origin of the divergence is the
$L$-dependence of the the parameter $b$ of Eq.(\ref{eq:tzeror}). Since the
phase
point $P_T^{\prime}$ is far enough from the triple point so that $b(L) \propto
\exp((\zeta_o(P_T^{\prime}) -\zeta_o(P_T))L^d)$, $r$ diverges as $r\propto
\exp((\zeta_o(P_T^{\prime}) -\zeta_o(P_T))L^d)$. Since circles of both sets
converge to the unit circles if we move $\eta_o$ from $\eta_o=0$, to a
coexistence point on either $B+$ or  $B-$ plane of Fig.~1, the two sets are
actually finite-size ghost circles that separate three phases bounded by  $B+$
and $B-$ plane.

Fig.~3C  is a plot of  zeros in $z_{\zeta}$-plane at $P_A$ and at two nearby
points, $P_A^{\pm}=(\zeta_o=\zeta_o(P_A) \pm 0.000005,\eta_o=\eta_o(P_A))$
Although zeros  for only two sizes, $L=8$ and $L=40$ are shown, zeros of all
sizes collapse into single curves. This  proves that $a=1$, which implies that
the partition functions of two phases are added up without a prefactor.

Fig.~3D is the plot of exact zeros at $P_A$, together with approximate zeros
calculated using Eqs.(\ref{eq:imaginaryzero}) and (\ref{eq:realzero}).  We
calculated $\hat{\gamma}_l$ for $l=1$ through $6$ and approximate zeros of
successive order which include $\hat{\gamma}_l$ of successively higher orders.
The reason that we see only $4$ sets of approximate zeros is because $a=1$ or
$\hat{R}=0$ makes the third and fifth terms of Eq.(\ref{eq:realzero}) vanish.
It
should be noted that $\hat{R}=0$  makes even-order terms of
Eq.(\ref{eq:imaginaryzero}) vanish instead. This again proves $a=1$.
It also shows that the locus of zeros for large $\theta$ is affected more and
more
by the discontinuities of higher order derivatives of the free energy,
$\tilde{\gamma}_l$'s as predicted by the theorem.

In conclusion we have shown that the extended circle theorem for first-order
transitions follows from a very simple property that the distribution function
for the order parameter is multiply peaked at the coexistence point.  In fact
zeros of the partition function can be expressed in terms of the position and
shape of these peaks through cumulants of each phase.  We have also found the
finite-size effect on the zeros of the partition function which can be useful
in
determining the order of transition as well as the accurate estimate of the
transition point from the numerical data of finite-size systems which can be
easily obtainable from a Monte Carlo technique of Ref.\cite{leek90} for
example.
Finally we have found that the partition functions are added up without a
prefactor in the case of asymmetric phase transition in the Blume-Capel model,
at least at low temperatures.

This work was supported in part by the Ministry of Education, Republic of Korea
through a grant to the Research Institute for Basic Sciences, Seoul National
University , in part by the Korea Science Foundation through Research Grant to
the Center for Theoretical Physics, Seoul National University and in part  by
S.N.U.  Daewoo Research Fund.  The author wishes to thank  Moo Young Choi and
Julian Lee
for the critical reading of the paper.

\begin{figure}
\caption{Schematic phase diagram of Blume-Capel model. The symmetric
coexistence
plane
$A$ lies in $T-\Delta$ plane bounded by  the triple line $OP$ and the critical
line $PR$.  Two asymmetric coexistence planes $B+$ and   $B-$ bifurcate from
the
triple line symmetrically. $P$ is the tricritical point.}
\label{fig1}
\end{figure}

\begin{figure}
\caption{$\ln(Z_o((N_++N_-,N_+-N_-))$ on ($N_+,N_-$) plane for $L=16$. I. At
$\zeta_o=-1, \eta_o=0$, a symmetric two-phase transition point corresponding to
Fig.~1(a).  II. At $P_T$.  III. At $\zeta_o=1, \eta_o=0$, a symmetric single
phase point far away from the triple point corresponding to Fig.~1(c).  IV. At
$P_A$.}
\label{fig2}
\end{figure}

\begin{figure}
\caption{Zeros of partition function.  Circles of a solid line are unit circles
in
all $4$ figures.  A. $6$ sets of $N(=L^2)$ zeros for the system size,
$L=8,10,12,14,16$ and $L=60$ in decreasing size of dots at $P_T$ in
$z_{\zeta}$-plane.  The radial distance from the unit circle is magnified 50
times.  Even at this magnification  zeros for $L=60$ is too close to the unit
circle to be discernible.  B.  Zeros $P_T^{\prime}$ in $z_{\eta}$-plane for
$L=8, 10,12$ and $L=14$.  Distance from the unit circle for outer circles is
scaled down by a factor of $1/25$ to bring them into the plotting area.  C.
Zeros in $z_{\zeta}$-plane at $P_A$(the circle tangent to the unit circle),
$P_A^-$(the one inside of the unit circle) and $P_A^+$(the largest circle which
crosses the unit circle).  In C  and D   the radial
distance from the unit circle are again magnified this time
$50000$ times!    D.  Exact and
approximate zeros at
$P_A$.  Zeros for two sizes, $L=8$ and $40$ are shown. The unit circle is the
zeroth order approximation, the innermost is the second order approximation,
zeros
closest to the unit circle is the fourth, and the sixth is closest to the exact
zeros, the circle in the middle.
 Similar plots for zeros
in $z_{\eta}$-plane (in the direction of Fig.~1(d)) are obtained at this
point.}
\label{fig3}
\end{figure}

\end{document}